\documentclass[aps,prl,preprint,groupedaddress,showpacs,showkeys]{revtex4-1}
\usepackage{color}
\usepackage{graphicx}
\usepackage{amsmath,amssymb}
\usepackage{amsfonts}
\usepackage{amsmath,amssymb}

\newcommand{\chem}[1]{\mbox{{$\rm #1$}}}

\newcommand{\ds}{\displaystyle}

\newcommand{\A}{\mbox{\rm {\AA}}}

\newcommand{\VEC}[1]{\mbox{\boldmath $#1$\unboldmath}}

\newcommand{\ket}[1]{\mbox{$\lvert{#1}\rangle$}}
\newcommand{\bra}[1]{\mbox{$\langle{#1}\rvert$}}
\newcommand{\braket}[2]{\mbox{$\langle{#1}\rvert{#2}\rangle$}}

\newcommand{\proA}{\mbox{\rm {\A}}\mbox{$^{-1}$}~}

\def\<={\raisebox{-0.5ex}{\small $\stackrel{<}{\sim}$}}
\def\>={\raisebox{-0.5ex}{\small $\stackrel{>}{\sim}$}}
\newcommand{\Eq}[1]{Eq.~({\ref{#1}})}
\newcommand{\Fig}[1]{Figure~{\ref{#1}}}

\newcommand{\kb}{\mbox{$k_{\rm B}$}}

\newcommand{\subi}[2]{\mbox{$ #1_{\rm #2}$}}

\newcommand{\spr}{^\prime}

\newcommand{\iu}{\chem{i}}

\newcommand{\abs}[1]{\lvert #1\rvert}

\newcommand{\Exp}[1]{{\rm e}^{\ds #1}}

\newcommand{\lP}{\left(}
\newcommand{\rP}{\right)}
\newcommand{\dxsq}{\delta_x^2}

\newcommand{\rav}[2]{\left\langle #1\right\rangle_{#2}}
\newcommand{\Tr}[1]{\chem{Tr}\left(#1\right)}
\newcommand{\rhotop}{\hat{\rho}^{(T)}}

\newcommand{\Hop}{\hat{H}}
\newcommand{\psit}{\psi^{(T)}_{\theta}}
\newcommand{\It}{I_{\theta}}
\newcommand{\Imod}{I_{\rm mod}}
\newcommand{\phik}[1]{\ket{\phi_{#1}}}
\newcommand{\qv}{\VEC{q}}
\newcommand{\kv}{\VEC{k}}
\newcommand{\pv}{\VEC{p}}
\newcommand{\xv}{\VEC{x}}
\newcommand{\kicked}[1]{\chi^{(T,\qv)}_{\theta,{\rm #1}}}
\newcommand{\dxsqc}{\tilde{\delta}_x^2}
\newcommand{\lT}{\subi{\lambda}{th}}
\newcommand{\tT}{\subi{\tau}{th}}
\begin{document}
\title{A quantum mechanical evaluation\\ of the intermediate scattering function}
\author{Oussama Bindech, Roberto Marquardt}
\email[corresponding author: ]{roberto.marquardt@unistra.fr}
\affiliation{Laboratoire de Chimie Quantique - Institut de Chimie\\
UMR~7177~CNRS/Universit\'e de Strasbourg \\
4, rue Blaise Pascal - CS 90032 - 
67081 STRASBOURG CEDEX - France}
\author{Fabien Gatti}
\affiliation{Institut de Sciences Mol\'eculaires d'Orsay, UMR
    8214 CNRS/Universit\'e Paris-Saclay, B\^at 520, Rue Andr\'e Rivi\`ere,
    91405 Orsay Cedex, France}
\author{Souvik Mandal, Jean Christophe Tremblay}
\affiliation{Laboratoire de Physique et Chimie Th\'eoriques,
    UMR 7019 CNRS/Universit\'e de Lorraine, 1 Blvd. Arago,
    57070 Metz, France }
\date{\today}
\begin{abstract}
The 
intermediate scattering function is interpreted as a correlation
function of thermal wave packets of the scattering centers perturbed
by the scattering particles at different times. A proof of concept is
given at the example of ballistic moving centers. The ensuing
numerical method is 
then illustrated at the example of CO adsorbed on Cu(100).
\end{abstract}
\maketitle
The van Hove formula~\cite{vanHove:1954} for the intermediate
scattering function (ISF) $I(\qv,t)$ 
has been used for more than half a century in the investigation of the
dynamics of condensed
matter~\cite[and references cited therein, this list being rather
  incomplete]{Vineyard:1957,Schofield:1960,ChudleyElliot:1961,Gunn:1984,Grabert:1989,Kob:1995,Ellis:2009,Townsend:2019}. It  
is given by the expression 
\begin{equation}
    I(\qv,t) = \sum\limits_{j}\;\sum\limits_{j\spr}\;
    \Tr{
      \rhotop
      \;\Exp{-\iu\qv\hat{\xv}_{j\spr}}
      \;\Exp{\iu\Hop t/\hbar}
      \;\Exp{\iu\qv\hat{\xv}_j}
      \;\Exp{-\iu\Hop t/\hbar}
    }\label{ISFdef}
\end{equation}
Here $\Hop$ is the Hamiltonian of the scattering centers at
positions $\xv_i$ and $\xv_j$,  $\qv=\Delta\VEC{p}/\hbar$ is the
wave vector pertaining to the change of momentum during the
scattering and $t$ is a time; $\rhotop=\Exp{-\Hop/(\kb T)}/Q$ is the
thermal density operator, where $T$ is the temperature of the
scattering system, and $Q$ is the canonical partition function;
$\hbar$ is the Planck constant divided by $2\pi$ and $\kb$ is the
Boltzmann constant. This function is also known as the space Fourier
transformation of the space-time pair correlation function. 

More recently, the ISF was also evoked in the context of helium
scattering experiments to determine diffusion 
coefficients of adsorbed atoms and molecules on
surfaces~\cite{Ellis:2009}. So far, in these investigations, the ISF
is mainly evaluated from molecular dynamics simulations following the laws of
classical mechanics, i.e. by interpreting the factors   
$
\Exp{\iu\Hop t/\hbar}
      \;\Exp{\iu\qv\hat{\xv}_j}
      \;\Exp{-\iu\Hop t/\hbar}
      =
      \Exp{\iu\qv\,\xv_j(t)}
      $
      in terms of the classical mechanical trajectories $\xv_j(t)$.

As was already noted by van Hove himself~\cite{vanHove:1954},
$I(\qv,t)$ cannot be cast into the form
$I(\qv,t)=\Tr{\rhotop\,\hat{\Upsilon}(t)}$
with some time dependent operator
$\hat{\Upsilon}(t) = \Exp{\iu\Hop
  t/\hbar}\;\hat{\Upsilon}(0)\;\Exp{-\iu\Hop t/\hbar}$ in the
Heisenberg representation. The ISF is thus not a conventional time
dependent quantum mechanical observable. Rather, it is a quantum 
correlation function~\cite{Kubo:1992}. While the classical evaluation
of the ISF relates to the correlation of particles at positions
$\xv(t)$ and $\xv(0)$ under thermal conditions, its pure quantum 
mechanical evaluation is not obviously connected to a time correlation of
positions, as the particles' density is constant and 
delocalized in a thermal quantum mechanical state. Townsend and Chin
derived an analytical quantum dynamical 
expression for the ISF on the basis of tight binding model
Hamiltonians and the Baker-Campbell-Hausdorff
disentangling theorem~\cite{Townsend:2019}, providing theoretical 
evidence for long-range coherent tunneling in helium-3 scattering
experiments. 

In this letter we demonstrate that the ISF can be interpreted as the 
correlation between two typical members of the thermal ensemble: the
first one is a thermal 
wave packet that is perturbed by the
interaction with the scattering particle beam at time $t_0$; the 
second one is a thermal wave packet perturbed at time $t_0+t$. A   
proof of principle is given at the example of the ballistic
particle, for which the ISF is known
analytically~\cite{vanHove:1954}. 
This interpretation allows us to evaluate the ISF from the quantum
dynamical evolution of the scattering centers in a general way. 
In order to illustrate the potential applicability of the ensuing
method, the ISF is then calculated within a model study of a CO
particle moving on a Cu(100) substrate.

Following Tolman~\cite{Tolman:1938}, the thermal density operator may
be considered to be a 
{statistically averaged} operator
$
\rhotop = \rav{\rhotop_{\theta}(t)}{\theta}
$
, where
$
  \rhotop_{\theta}(t) =
  \ket{\psit(t)}\bra{\psit(t)}
$
.
  The states $\ket{\psit(t)}$ constitute a stochastic ensemble of
  time dependent, pure states
 characterized by random
variables $\theta$ and called 
\textit{stochastic thermal wave packets}:
\begin{eqnarray}
\ket{\psit(t)} &=&
\sum\limits_n\,\frac{\Exp{-E_n/(2\kb T)+\iu\,\theta_n-\iu\,E_n\,t/\hbar}}{\sqrt{Q}}\,\phik{n}\label{TWP}
\end{eqnarray}
Here, it is assumed that the system of scattering particles has a
discrete spectrum with system's eigenstates $\phik{n}$:
$\Hop\phik{n}=E_n\phik{n}$. Note that the statistical average
annihilates the time dependency of $\rhotop_{\theta}(t)$, so that one
may as well write $\rhotop = \rav{\rhotop_{\theta}(0)}{\theta}$.  

For simplicity, and without lack of generality,
consider in the following a system composed of a
single particle. Because taking the ensemble average
$\rav{\cdot}{\theta}$ and the trace $\Tr{\cdot}$ commute, we may set
\begin{eqnarray}
  I(\qv,t)
  &=& 
\Tr{\rav{\rhotop(0;\theta)}{\theta}\;\Exp{-\iu\qv\hat{\xv}}\;\Exp{\iu\Hop t/\hbar}\;\Exp{\iu\qv\hat{\xv}}\;\Exp{-\iu\Hop
    t/\hbar}} = \Big\langle\It(\qv,t)\Big\rangle_{\theta}\label{ISFens}
\end{eqnarray}
where
\begin{eqnarray}
  \It(\qv,t)
  &=&
\Tr{\ket{\psit(0)}\bra{\psit(0)}\;\Exp{-\iu\qv\hat{\xv}}\;\Exp{\iu\Hop t/\hbar}\;\Exp{\iu\qv\hat{\xv}}\;\Exp{-\iu\Hop t/\hbar}}\nonumber\\[5mm]
  &=& 
\bra{\psit(0)}\;\Exp{-\iu\qv\hat{\xv}}\;\Exp{\iu\Hop
  t/\hbar}\;\Exp{\iu\qv\hat{\xv}}\;\Exp{-\iu\Hop t/\hbar}\;\ket{\psit(0)}
\end{eqnarray}

Note that $\Exp{-\iu\Hop t/\hbar}\;\ket{\psit(0)} = \ket{\psit(t)}$. Let now 
\begin{equation}
\ket{\kicked{KE}(0)} = 
\Exp{\iu\qv\hat{\xv}}\;\ket{\psit(0)}
\end{equation}
be a thermal state that is 
\textit{``kicked''} by the \textit{scattering operator}
$\Exp{\iu\qv\hat{\xv}}$ at time zero. Indeed, the state
$\ket{\chi^{(\kv,\qv)}} = \Exp{\iu\qv\hat{\xv}}\ket{\kv}$ resulting
from the action of the scattering operator on a momentum eigenstate
$\ket{\kv}$ with $\hat{\pv}\ket{\kv}=\hbar\kv\ket{\kv}$ yields just
the shifted momentum eigenstate $\ket{\kv+\qv}$. 
%

Its time evolution is then
$\ket{\kicked{KE}(t)} = \Exp{-\iu\Hop
  t/\hbar}\;\ket{\kicked{KE}(0)}$. Similarly, let 
\begin{equation}
\ket{\kicked{EK}(t)} = 
\Exp{\iu\qv\hat{\xv}}\;\ket{\psit(t)}
\end{equation}
be the \textit{same} thermal state that has evolved with time and is then 
``kicked'' by the \textit{same} scattering operator.

We may then write
\begin{eqnarray}
  \It(\qv,t) &=&
\bra{\kicked{KE}(0)}\;\Exp{\iu\Hop t/\hbar}\;\Exp{\iu\qv\hat{\xv}}\;\Exp{-\iu\Hop
  t/\hbar}\;\ket{\psit(0)}\nonumber\\
  &=& 
\bra{\kicked{KE}(t)}\;\Exp{\iu\qv\hat{\xv}}\;\ket{\psit(t)}\nonumber\\
  &=& 
\braket{\;\kicked{KE}(t)\;}{\;\kicked{EK}(t)\;}\label{ISFTWP}
\end{eqnarray}

Individual statistically tagged ISF functions $\It(\qv,t)$ can thus be
understood as yielding the \underline{correlation}
between a kicked and then time evolved 
  thermal wave 
packet $\chi_{\rm KE}$ and \underline{the same} time evolved and then kicked thermal
wave packet $\chi_{\rm EK}$. The latter represent two states of the
scattering centers that result from
their interaction with the helium atoms at different times. The observable ISF is then obtained as the
ensemble average over these correlations in~\Eq{ISFens}. Such
correlations sample the degree of order in the statistical
ensemble. At finite temperature, they should decay in time as the
delay between the two events, i.e. the ``kicks'', increases. This is
precisely the observed behavior for the experimental ISF, although 
experimentally, other processes might also contribute to the decay. 

As a proof of concept, we evaluate~\Eq{ISFTWP} numerically for a free
particle of mass $m$ moving at constant temperature. In this case the
ISF is known analytically~\cite{vanHove:1954}:
\begin{eqnarray}
I(q,t) 
    &=& \Exp{-\frac{\dxsqc(t)\,q^2}{2}}
\end{eqnarray}
where 
%
$\dxsqc(t) = \frac{\ds \kb T}{\ds m}\,t^2 - \iu\,\frac{\ds \hbar}{\ds m}\,t$ 
is a complex valued \textit{mean square displacement} (MSD). Note that
$\Re[\dxsqc(t)]$ is the classical mechanical result for the MSD of the ballistic particle. 
In the
natural units 
$\lT = h/\sqrt{2\pi\,m\,\kb\,T}$ and $\tT = \hbar/(\kb\,T)$, the MSD
becomes $\dxsqc(t) = \lT^2/(2\pi)\times\lP\,(t/\tT)^2 - \iu\,
t/\tT\,\rP$, so that
\begin{equation}
  \begin{array}{rcl}
  -\ln\lP\abs{I(q,t)}\rP &=& \frac{\ds \lP\lT\,q\rP^2}{\ds
    4\pi}\;\lP\frac{\ds t}{\ds \tT}\rP^2\\
  \arg\lP I(q,t)\rP &=& \frac{\ds \lP\lT\,q\rP^2}{\ds 4\pi}\;\frac{\ds
    t}{\ds \tT}
  \end{array}\label{ISFbal}
\end{equation}

\begin{figure}[h!]
 \begin{center} \includegraphics[width=12cm]{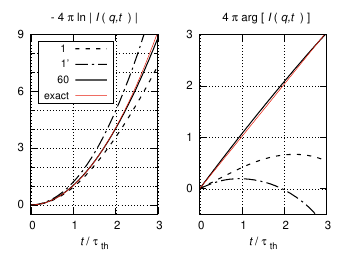}
    \parbox{14cm}{
      \caption{\label{FPISF} Quantum dynamical time evolution 
        of the ISF of a ballistic particle
        ($T=300\,\chem{K},m=1\,\chem{u}$) at $q=1\,\proA$. The left hand
        side relates to the amplitude, the right hand side to the
        phase of the ISF. Black lines relate to $\It(q,t)$ for individual stochastic
        samples (\Eq{ISFTWP}, dotted), and for an ensemble average
        (\Eq{ISFens}, continuous). Red lines give the exact results
        from~\Eq{ISFbal}. 
}
    }
    \end{center}
\end{figure}

\Fig{FPISF} shows the numerical evaluation of these functions (black
continuous and dotted lines) and the exact values (red
lines) for $\lT\,q = 1.007\,951$ ($q=1\proA, T=300\,\chem{K},
m=1\,\chem{u}$).
Here, the Schr\"odinger equation is solved for the 
free particle Hamiltonian in a box of length $20\,\lT$ on a grid of
800 points. For each stochastic sample, calculations are converged to
within $10^{-3}$ relative precision 
in length and time over the time interval presented in the
figure. Numerical results are surprisingly smooth. Yet, 
single stochastic samples $\It(q,t)$ differ significantly from the
exact values as time evolves. Only after some substantial averaging,
60 in the example, one starts to achieve acceptable agreement between the
numerical and the analytical results. Interestingly, the phase of the ISF
needs somewhat more samplings to converge than its amplitude.

Evaluating Eqs.~(\ref{ISFTWP}) and then (\ref{ISFens}) is a method to 
determine the ISF from first principle calculations. 
As an illustration, we evaluate these 
formulae for a simplified, albeit realistic model of CO moving on
Cu(100). This system was investigated in helium-3
experiments~\cite{Ellis:2004,Ellis:2009b}.     
Here, however, we are not attempting to calculate accurately experimental 
rates. Rather, we wish to show that the evaluation of~\Eq{ISFTWP} leads to  
a qualitative temporal behavior of the ISF 
that is also found in experiments at realistic time scales. 

\clearpage

We consider a 
one-dimensional model of the system along the $\langle 1 0 0\rangle$ 
crystallographic direction. The periodic model potential energy
function was 
developed in refs.~\citenum{roma:45} and~\citenum{roma:68} from first
principle calculations. The model
potential was also used
in previous related work~\cite{roma:79,roma:97}, were defining parameters are
given. As reported in ref.~\citenum{roma:68}, the barrier energy for jumps between 
stable CO adsorption sites is 33.5~meV and includes a significant
variation of the zero point energy. The potential accommodates 13
bands of 
hindered translational levels up to this barrier, which
correspond classically to confined vibrational motion within an adsorption
well. The first excited band lies about 1.8~meV above the ground
level. With increasing energy, bands become broader due to
tunneling to neighboring wells~\cite{roma:79}.   
At 190~K, for instance, the higher most band below
the barrier still has a relative population of about 15\% with respect to the
ground level. The thermal wave packet contains thus a large portion of
energy eigenstates below the barrier, but also states above it. 
In the numerical evaluation, a grid of 80
elementary cells with individual cell length of $2.556\,\A$, the
Cu(100) lattice constant, and
50 points per cell was used, which yields  
converged calculations. Ensemble averages are converged with 20
samples. 

\begin{figure}[h!]
 \begin{center} \includegraphics[width=12cm]{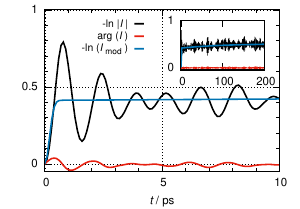}
    \parbox{14cm}{
      \caption{\label{COISF} Quantum dynamical time evolution 
        of the ISF of CO ($m=27.9949\,\chem{u}$) moving along the
        $\langle 1 0 0\rangle$ 
        direction on the Cu(100) crystal surface at 190~K and
        $q=1\,\proA$ following \Eq{ISFens} with 20 samples. Black
        line: 
        $-\ln\lP\abs{I(q,t)}\rP$. Red line:
        $\arg\left[I(q,t)\right]$. Blue line: $-\ln\lP\Imod(t)\rP$
        as defined in~\Eq{ISFmod}.     
}
    }
    \end{center}
\end{figure}

Initially, $-\ln\abs{I(q,t)}\sim t^2$ (black line), followed by
piece-wise periodic 
oscillations with smaller amplitudes around a median, slowly ascending
line (drawn in blue color)  
and periods of about 1.5~ps. The phase of the ISF (red line)
oscillates in a similar irregular way around zero.   
Note that, $-\ln(I(q,t))\propto \dxsq(t)$, the mean square
displacement
(MSD)~\cite{Vineyard:1957,Schofield:1960,ChudleyElliot:1961}. 
A typical behavior
found in early experimental work on neutron scattering is explained by
the jump model of Chudley and Elliott, which ``predicts a width like
the Debye model with some slight increase at large
$t$"~\cite{ChudleyElliot:1961}. This behavior of the ISF is quite
generally also
found in the spin echo polarization decay and is featured by the black
line 
in~\Fig{COISF}. 

The MSD 
of a harmonic oscillator behaves as $\dxsq(t)\propto \sin^2(\omega
t)$ also quantum mechanically~\cite{roma:97} which explains the
initial quadratic increase of the black line. The approximate period
of 1.5~ps matches roughly $\tau = h/(2\Delta E) \approx
1.14\,\chem{ps}$ from the fundamental transition energy $\Delta E
\approx 1.8\,\chem{meV}$. The potential is highly
anharmonic, though, which explains the irregular periodic
behavior and the imperfect matching of periods. Quite generally, the
quantum dynamics of  
anharmonic vibrations becomes rapidly irregular~\cite{roma:8}. This apparent
lack of coherence is even more pronounced for strongly anharmonically
coupled vibrations in high dimensional systems where survival
probabilities decay approximately exponentially and oscillations are
``coherently'' damped~\cite{roma:11}. In the context of apparently
statistical time-dependent quantum dynamics we should mention the
famous Bixon-Jortner model for radiationless
transitions~\cite{Jortner:68}, see also ref.~\citenum{roma:87} and references
  cited therein.   

In the present, one-dimensional study, oscillations persist for longer
times, as can be seen from the inset. The 
blue median line around which the ISF oscillates results from a model, 
smooth
representation of the ISF.  
In the analysis of the experimental 
work, the ISF is often modeled by a 
double-exponential decaying function~\cite{Ellis:2009b} which
smooths down the oscillations that are observed in the initial phase of the
ISF evolution. Here, we use
the model function
\begin{equation}
  \Imod(t) = 1 + P_1\,\lP\Exp{-A_1\,t^2\,+\,\lP P_2\,A_2\,/\,P_1\rP\,t}-1\rP + P_2\,\lP\Exp{-A_2\,t}-1\rP\label{ISFmod}
\end{equation}

This model ensures that $-\ln\lP\Imod(t)\rP \propto t^2$ for short
times, which is a reasonable assumption to make, as for short times
the ballistic character of the motion prevails, whether the particle
is moving freely or in a harmonic 
potential well. 
Adjustment of $\Imod$ to the data defining the black line up to 200~ps
yields  
$
 P_1 = (0.338\pm 0.003),
 A_1 = (15\pm 4)\times\chem{ps}^{-2}, 
 P_2 = (0.059\pm 0.005),
 A_2 = (1.1\pm 0.3)\times 10^{-3}\,\chem{ps}^{-1}
 $
 (uncertainties are 68\% confidence intervals) which define the blue line
 in~\Fig{COISF}. They hold for $q=1\,\proA$. To test their 
 reliability, the 
 ISF was calculated 
 to up to 400~ps, while model and parameters keep on describing well the median of the
 extended function.
We note that each realization $\It$ from 
\Eq{ISFTWP} is
the result of a coherent quantum 
dynamical evolution of a thermal wave packet. The  
stochastic nature of the latter, which stems from the random phases
in~\Eq{TWP}, induces some fluctuation between individual realizations. 
The ISF depicted in~\Fig{COISF} was obtained from   
averaging 20 individual stochastic realizations of the ISF, \Eq{ISFens}. 

%

The time Fourier transformation of the ISF yields the dynamical
structure factor (DSF)~\cite{vanHove:1954}. The first exponential
in~\Eq{ISFmod} 
gives a broad Gaussian, the second a much sharper Lorentzian DSF. In
the analysis of the spin-echo data, one usually
considers the quasielastic broadening of the sharper part of the DSF as
indicative of intercell diffusion~\cite{Ellis:2004b}. The
corresponding rate of the ISF is called 
 \textit{dephasing rate} in spin-echo 
experiments~\cite{Ellis:2009b}. It is interesting to
compare the value of $A_2$ with  
dephasing rates. 

 The time scale of the slow decay rate $A_2$ is indeed comparable 
 with the dephasing rate $\alpha 
 \approx 0.76\times 10^{-3}\,\chem{ps}^{-1}$ from figure 7 of  
 ref.~\citenum{Ellis:2009b} for $q=1\,\proA$ in the
 $\langle 1 0 0\rangle$ direction. 
 This is nearly 3/4 of $A_2$. Before drawing any further conclusion
 from this comparison with respect to the quality of the treatment
 and, in particular, to the accuracy of the  
 model used, one has to bear in mind the
 major approximations   
 made to obtain~\Fig{COISF}. Caveats of the present  
 model are the one-dimensional treatment and the neglect of  
  dissipation during the dynamics. Furthermore, the potential energy
 function model might be incorrect, despite its origin from first
 principle calculations. The rough  
 agreement between $A_2$ and $\alpha$ could just be a fortuitous
 compensation of errors. 

 \Fig{COISF} shows nevertheless that~\Eq{ISFTWP} is tenable to pick up
 a temporal behavior of the ISF that is found experimentally at
 comparable time scales. Intrinsic quantum dynamical properties of 
 the motion such as the non-local distribution of thermal probability
 densities, irregular interference effects between waves pertaining to
 individual eigenstates of the system in the thermal wave packet and, last but not least,
 tunneling quite naturally lead to a roughly exponential decay  
 of the ISF in the long time limit that is comparable with
 experimental findings despite the neglect of explicit dissipative effects in the
 quantum dynamics.  
 To the best of our knowledge, such a behavior has
 so far been exclusively rationalized in  
 terms of dissipative dynamics. 
 
 The present theoretical 
 treatment can be extended in order to include the multidimensional
 character of the 
 motion as well as possible channels of  
  dissipation during the dynamics~\cite{roma:93,roma:98}. Dissipation
  has been considered insofar as 
  the thermal wave packet,
\Eq{TWP}, is the result of friction and its random
  phases 
  mirror the resulting fluctuations. Dissipation is a continuous
  random process, however, and the 
  time evolution of the thermal wave packet in \Eq{TWP} lacks the ongoing influence of
  the 
  latter. For CO/Cu(100), friction is expected to become relevant in
  time intervals beyond 8~ps~\cite{Ellis:2004,Ellis:2004b}. One could
  therefore expect the 
  ISF depicted in 
  the main part of~\Fig{COISF} to be essentially invariant in the presence of
  friction. For the inset up to 200~fs and beyond we presently cannot
  foresee its effect on the ISF in~\Eq{ISFTWP}. One should expect that 
  friction does not alter the 
absolute values of the coefficients in the thermal wave packets; very 
likely, friction could merely lead to a random reshuffling of their 
phases at random times. In the analysis of the experimental data,
friction is partially responsible for the vibrational  
dephasing and the consequent reduction of the oscillatory behavior of
the ISF~\cite{Ellis:2004b}. We note that the latter might as well
result from ``coherent'' damping in   
multidimensional vibrational or vibronic 
dynamics~\cite{roma:11,roma:87}.  

Beyond being a rather straightforward method to 
  evaluate the ISF within the framework of quantum mechanics, 
 \Eq{ISFTWP} is in the first place a novel  
 interpretation of this quantity as a quantum time correlation 
 function between two states, namely a perturbed (kicked) then 
 time evolved 
 and a time evolved then perturbed (kicked) 
 thermal wave packet. These are states of the scattering centers at
 thermal conditions having interacted with the scattering beam at
 different times. Individually, thermal wave packets are pure
 states and only ensemble averages are reasonably comparable with
 experimental findings obtained under these conditions.
Further work to explore the $\qv$ and $T$-dependence of the ISF is in
progress. It will in particular be aimed at  
 including friction and higher dimensions in the dynamical
 calculations to ensure the predictability and accuracy of the 
method.

\begin{acknowledgments}
This work benefits from a grant received from the French Agence
Nationale de la Recherche (ANR) under project QDDA. We wish to express
our gratitude to Hans-Dieter Meyer and all the developers of the
Heidelberg MCTDH program package, which was used for the numerical
work presented here. 
\end{acknowledgments}
\newcommand{\Aa}[0]{Aa}
%
\end{document}